\RequirePackage{lineno}
\documentclass[10pt,numberedappendix]{emulateapj}

\usepackage{graphicx}% Include figure files
\usepackage{dcolumn}% Align table columns on decimal point
\usepackage{bm}% bold math
\usepackage[nolist,nohyperlinks]{acronym}
\usepackage{hyperref}% add hypertext capabilities
\usepackage{color}
\usepackage{xspace}
\usepackage{subfigure}
\usepackage[dvipsnames]{xcolor}
\usepackage{amsmath}
\usepackage[normalem]{ulem} %for \sout

\usepackage{hyperref}

\newcommand{\err}{\ensuremath{\mathcal{R}}\xspace}

\newcommand{\prlsec}[1]{\textit{#1} -- \xspace}

\newcommand{\LIGOlabMIT}{\affiliation{LIGO Laboratory, Massachusetts Institute of Technology, 185 Albany St, Cambridge, MA 02139, USA}}
\newcommand{\MKI}{\affiliation{Department of Physics and Kavli Institute for Astrophysics and Space Research, Massachusetts Institute of Technology, 77 Massachusetts Ave, Cambridge, MA 02139, USA}}
\newcommand{\UNH}{\affiliation{Department of Physics \& Astronomy, University of New Hampshire, 9 Library Way, Durham NH 03824, USA}}
\begin{document}

\title{{The relative contribution to heavy metals production from binary neutron star mergers and neutron star-black hole mergers}}

\author{Hsin-Yu Chen}
\email[]{himjiu@mit.edu}
\thanks{NHFP Einstein fellow}
\author{Salvatore Vitale}
\email[]{salvo@mit.edu}
\LIGOlabMIT
\MKI
\author{Francois Foucart}
\email[]{Francois.Foucart@unh.edu}
\UNH

\begin{abstract}
The origin of the heavy elements in the Universe is not fully determined. Neutron star-black hole (NSBH) and {binary neutron star} (BNS) mergers may both produce heavy elements via rapid neutron-capture (r-process). 
We use the recent detection of gravitational waves from NSBHs, improved measurements of the neutron star equation-of-state, and the most modern numerical simulations of ejected material from binary collisions to measure the relative contribution of NSBHs and BNSs to the production of heavy elements.
As the amount of r-process ejecta depends on the mass and spin distribution of the compact objects, as well as on the equation-of-state of the neutron stars, we consider various models for these quantities, informed by gravitational-wave and pulsar data. We find that in most scenarios, BNSs have produced more r-process elements than NSBHs over the past 2.5 billion years. If black holes have preferentially small spins, BNSs can produce at least twice of the amount of r-process elements than NSBHs. If black hole spins are small and there is a dearth of low mass ($<5M_{\odot}$) black holes within NSBH binaries, BNSs can account for the near totality of the r-process elements from binaries. For NSBH to produce large fraction of r-process elements, black holes in NSBHs must have small masses and large aligned spins, which is disfavored by current data.

\end{abstract}

\maketitle
\section{Introduction} 
Neutron star-black hole (NSBH) and binary neutron star (BNS) mergers are among possible astrophysical formation sites of heavy elements through rapid neutron-capture (r-process) nucleosynthesis{~\citep{1974ApJ...192L.145L}}. The tidal disruption of a neutron star (NS) by its black hole (BH) companion, and the subsequent ejection of neutron-rich material into the interstellar medium was originally proposed by{~\citet{1974ApJ...192L.145L,1976ApJ...210..549L}}. Their predictions have largely been confirmed by modern numerical simulations: as long as the BH is of sufficiently low mass for tidal disruption to occur, $\sim (0.01-0.1)M_\odot$ of neutron-rich material can be ejected during a NSBH merger~\citep{Foucart:2013a,kyutoku:2015}. BNS mergers typically do not lead to as much mass ejection ($\lesssim 0.01M_\odot$;~\citet{hotokezaka:13,Dietrich:2016fpt}), unless their mass ratios are highly unequal (possibly $\lesssim 0.03M_\odot$ for systems with NS masses $\gtrsim 1.1M_\odot$;~\citet{Dietrich:2016hky,Kiuchi:2019lls}). On the other hand, disrupting NSBH systems and most BNS systems can both produce compact remnants surrounded by massive accretion disks, with a significant fraction of these disks expected to become unbound within a few seconds of the merger~\citep{Fernandez2013,Siegel:2017nub,Christie:2019lim}. Roughly $\sim 0.01M_{\odot}$ of mass can be ejected during the post-merger evolution. 

The relative importance of BNSs, NSBHs, and other potential sources of r-process elements (e.g. collapsars;~\citet{Surman:2005kf,Siegel:2018zxq}), whose potential as sources of r-process elements is still under debate~\citep{Miller:2019mfl,Fujibayashi:2020jfr}, and magnetorotational core-collapse supernovae(~\citet{2012ApJ...750L..22W,Mosta:2017geb,2021Natur.595..223Y}, which however cannot easily produce the heavier r-process elements) remains highly uncertain. 
To determine the relative contribution of BNSs and NSBHs to the production of r-process elements one needs to quantify a) how much r-process material is ejected by each system as a function of its parameters and b) the merger rate of BNS and NSBH as a function of the system's parameters. Predictions might be compared to measurements of r-process abundances on Earth~\citep{Paul:2001fm,2021Sci...372..742W}, in the solar system~\citep{1993Metic..28Q.399M,2021Sci...371..945C}, and in stars other than the Sun~\citep{2012ApJ...750...76R,2016Natur.531..610J,FREBEL2019167909,2020ApJS..249...30H}. 

The observation of a kilonova following the first gravitational wave (GW) detection of a BNS, GW170817~\citep{2017PhRvL.119p1101A,GBM:2017lvd,2017Sci...358.1556C}, is consistent with the models for an emission powered by radioactive decays of heavy elements produced through r-process nucleosynthesis (see e.g.~\citet{Metzger:2019zeh} for a review) with $\lesssim 0.05M_\odot$ of ejected matter. Significant uncertainties exist arising from the details of the nuclear physics processes, as well as the composition and complex 3D geometry of the outflows. The observations of ancient dwarf galaxies also favor r-process enrichment from rare events, such as BNSs, producing copious amounts of r-process material~\citep{2016Natur.531..610J}. It thus seems very likely that {\it some} r-process nuclei are produced in BNSs. Indeed, using the astrophysical rate of BNS mergers inferred from the GW observations, one can estimate the BNS contribution to the production of r-process elements~\citep{2010MNRAS.406.2650M,LIGOScientific:2017pwl}. 
On the other hand, the recent discovery of two NSBHs by the LIGO-Virgo-Kagra (LVK) collaboration, GW200105 and GW200115~\citep{nsbhdiscovery}, has provided the first direct evidence of this type of systems. 
The merger rate and intrinsic properties inferred from these new sources enable novel constraints on the relative contribution to the production of r-process elements from BNSs and NSBHs.

\section{Methods} The amount of mass ejecta from binary mergers, and thus the r-process yield, depends on the mass and spin distribution of compact objects, as well as on the equation-of-state (EoS) of NSs.
As none of these are exactly known, we consider several possible models, which are consistent with GW and pulsar measurements. For each model, we generate populations of BNSs and NSBHs, use fits to numerical simulations to assess the amount of mass ejecta from each binary, and calculate the total contribution from BNSs and NSBHs. Each step of the analysis is described in turn here below.

\subsection{Simulated populations} 
We consider several distributions of mass and spin for BNSs and NSBHs. For the NSs (in both BNSs and NSBHs) we use the mass distribution of ~\citet{2020RNAAS...4...65F}, in which the observed Galactic NS masses were modelled as a bimodal distribution (See also~\citet{2018MNRAS.478.1377A}). We further restrict the NS mass to the range $1M_{\odot}<m_{\rm NS}<m_{\rm TOV}$, where $m_{\rm TOV}$ is the EoS dependent maximum mass of a cold and non-rotating NS. 
Recent work has also shown that a mixture of two Gaussians 
fits well the NSs detected in LVK data~\citep{landryread}. 
For the BHs, we consider three different mass distributions: a) uniform-in-log in the mass range $[5,40] M_{\odot}$, b) uniform-in-log in the mass range $[m_{\rm TOV},40] M_{\odot}$, and c) the \texttt{Power Law +Peak with GW190814} mass distribution from ~\citet{o3apop}. 
The models a) and b) consider a generic BH mass function with and without the observational mass gap between NSs and  BHs~\citep{2012ApJ...757...36K}. 
The scenario c) adopts the primary BH mass distribution of LVK binary black hole mergers (BBHs) assuming the secondary component mass of GW190814 is a BH~\citep{o3apop}. Not only does this model excludes the existence of a mass gap~\citep{2012ApJ...757...36K}, but it also allows for more BHs with masses in the range $\sim 2-3M_{\odot}$, potentially generated by BNS mergers~\citep{2021MNRAS.500.1817L,2021MNRAS.502.2049L}. 

We treat the NS as non-spinning. For the BH spins, we consider two possibilities: a) dimensionless spin magnitudes uniformly distributed in the range $[0,0.95]$ and aligned with the binary total angular momentum; b)  spin magnitudes and tilt angles that follow the primary BH spin distribution reported by the LVK in~\citet{o3apop}. 

In Table~\ref{tab:ratio} we list all six mass and spin models. For each model, we generated 100,000 simulated BNSs and NSBHs and calculate the ejected mass as described below.

\begin{table*}[]
    \centering
    \begin{tabular}{|c|c|c|c|c|}
    \hline
    Label &$m_1$  & $|\chi_1|$ & Tilt & $M_{\rm{ej,NSBH}}/M_{\rm{ej,Total}}$ \\
    \hline
    \texttt{Gap+aligned spin} &Uniform in log, $[5,40] M_{\odot}$ &  Uniform in [0,0.95] & Aligned &30\%  \\
    \texttt{Gap+BBH-like spin} &Uniform in log, $[5,40] M_{\odot}$ &  BBH-like & BBH-like &1\%  \\
    \texttt{No gap+aligned spin} &Uniform in log, $[m_{\rm TOV},40] M_{\odot}$  & Uniform in [0,0.95] & Aligned    & 49\% \\
    \texttt{No gap+aligned spin} &Uniform in log, $[m_{\rm TOV},40] M_{\odot}$  & BBH-like & BBH-like    & 11\% \\
    \texttt{BBH-like mass+aligned spin} &BBH-like &  Uniform in [0,0.95] & Aligned &77\%  \\
    \texttt{BBH-like mass+spin} &BBH-like & BBH-like & BBH-like  &35\%  \\
    \hline

    \end{tabular}
    \caption{Summary of BH mass and spin models explored in this paper. For the NSs in both BNSs and NSBHs, we use a bimodal distribution
fitted to the Galactic NS population~\citep{2018MNRAS.478.1377A,2020RNAAS...4...65F}. The label \texttt{BBH-like} represents the primary BH distribution inferred from the LVK binary black hole merger (BBH) observations~\citep{o3apop}, including GW190814. We stress that this model predicts more BHs with masses in the range $\sim 2-3M_{\odot}$ and small spin-aligned magnitudes. The last column reports the highest possible NSBH mass ejecta fraction given the 90\% upper limit of the NSBH/BNS astrophysical rate ratio. 
}
    \label{tab:ratio}
\end{table*}

\subsection{Estimation of ejected mass}
BNSs and NSBHs eject mass via the tidal disruption of NSs, the disk outflows in the post-merger remnant phase, and, for BNSs, during the collision of the two NSs. The post-merger outflows themselves can be further subdivided into early outflows from spiral arms in the remnant~\citep{Vsevolod:2020pak}, magnetically-driven winds~\citep{Siegel:2017nub,Christie:2019lim}, neutrino-driven winds~\citep{2015MNRAS.448..541J}, and thermal outflows in the advection-dominated disk formed late in the evolution of the remnant~\citep{Fernandez2013}. Each outflow component may have different composition, temperature, and velocity, impacting both the outcome of r-process nucleosynthesis~\citep{Wanajo2014,Lippuner2015} and the properties of the associated kilonova~\citep{Kasen:2013xka}. While a reliable model of all of the outflow components is not currently available, analytical fits to numerical simulations of BNSs and NSBHs exist. These provide estimates for the total amount of mass of the dynamical ejecta (mass ejected during the first few milliseconds following the merger) and of post-merger disks, as well as for the fraction of the disk that will be unbound after the merger. As we focus on the total mass of matter unbound by merger events, these estimates will be sufficient for our purpose.

We model the total mass of ejecta from a given binary as:
\begin{equation}\label{eq:tot}
    m_{\rm{ej}}=\alpha_{\rm{dyn}}m_{\rm{dyn}}+f_{\rm{loss}}m_{\rm{disk}}
\end{equation}
where $m_{\rm{dyn}}$ represents the mass of the dynamical ejecta, $m_{\rm{disk}}$ the mass of the disk formed in the post-merger phase, and $f_{\rm{loss}}$ the fraction of mass ejected from the disk. 
We also introduce a scaling factor $\alpha_{\rm{dyn}}$ that will be varied to account for uncertainties in the knowledge of $m_{\rm{dyn}}$. 

In order to estimate $m_{\rm{dyn}}$ and $m_{\rm{disk}}$, we use analytical fits to numerical simulations. For BNSs, we use Equation 6 of ~\citet{dyn} for the dynamical ejecta and Equation 4 of ~\citet{dyn} for the disk. For NSBHs, we use Equation 9 of ~\citet{dyn} for the dynamical ejecta and Equation 4 of ~\citet{rem} for the disk~\footnote{We note that the original formula for the disk only takes into account the BH spin magnitude (not its orientation) due to the spin-aligned numerical results. Therefore, we replace the spin magnitude with the component of the spin along the orbital angular momentum when dealing with tilted BH spins}. When the baryonic mass of NSs is needed in these formulae, we estimate it from the gravitational mass of the NS using Equation 33 of ~\citet{baryon}. 

The analytical formulae for NSBH binaries are valid for the range of EoS used in this study, as long as the aligned component of the dimensionless BH spin is $\chi_{\rm BH}\lesssim 0.9$. In these cases, we expect $\sim (10-20)\%$ accuracy~\citep{dyn,rem}. On the other hand, there are no simulations for very large mass ratios ($M_{\rm BH}/M_{\rm NS}\gtrsim 8$) or very compact stars. However, this is not a serious limitation since it's known that the NS is not disrupted in these cases (except for extreme BH spin magnitudes). The analytical formulae we use correctly predict a lack of mass ejection for large mass ratios.
For BNS mergers, uncertainties are typically larger. Even in the regions of parameter space where numerical simulations are available, $(30-50)\%$ errors in the fitting formulae are to be expected~\citep{dyn}. As a result, we choose a range of $\alpha_{\rm dyn}$ that covers the expected uncertainty in $m_{\rm dyn}$, i.e. $\alpha_{\rm dyn,BNS}\in [0.5,1.5]$ for BNSs and $\alpha_{\rm dyn,NSBH}\in [0.8,1.2]$ for NSBHs. Additionally, not enough simulations are available for asymmetric BNSs ($m_1/m_2\gtrsim 1.3$) to reliably predict the amount of mass ejection in that regime. No numerical simulation involving NSs of mass $\geq 1.1M_\odot$ has found dynamical mass ejection $>0.03M_\odot$, while extrapolation of the fitting formulae towards asymmetric binaries would lead to much more ejecta ($\gtrsim 0.06M_\odot$). To avoid issues with a small number of asymmetric binaries ($<15\%$ in our simulations) creating unphysically large amounts of r-process elements, we impose a cap $m_{\rm cap}=0.03M_\odot$ on the dynamical ejecta for BNSs.

We vary $f_{\rm loss}$ to account for the uncertainty in the physics of post-merger remnant and in the value of $m_{\rm disk}$. Recent studies of BH-disk systems indicate that $(5-20)\%$ of the mass of the disk will be ejected in magnetically-driven winds~\citep{Siegel:2018zxq,Christie:2019lim}, a comparable amount will be ejected through thermal outflows on longer timescales~\citep{2020MNRAS.497.3221F}, and additional mass ejection is expected during the circularization of the accretion disk~\citep{Kiuchi:2015qua}. In NS-disk systems, additional mass ejection is possible due to spiral-arm instabilities in the disk~\citep{Vsevolod:2020pak}, and most of the disk may be unbound in the presence of a long-lived NS remnant~\citep{Metzger:2014ila,Fahlman:2018llv}. 
Therefore we expect higher $f_{\rm loss,BNS}$ if a long-lived NS-disk system is formed after the BNS merger instead of a promptly collapsed BH-disk system~\citep{2014MNRAS.441.3444M}. 
On the other hand, even if the BNS collapses into a BH shortly after the merger, the resulting BH-disk systems would have lower mass than the systems formed from NSBH mergers and are expected to yield higher $f_{\rm loss}$~\citep{2020MNRAS.497.3221F}. 
Accordingly, we take $f_{\rm loss,NSBH}\in [0.15,0.60]$ for NSBH systems (which always form BH-disk systems), $f_{\rm loss,BNS}\in [0.15,1.0]$ for BNS systems (which may form BH-disk or NS-disk systems depending on when/if the remnant collapses to a BH), and we assume $f_{\rm loss,BNS}\geq f_{\rm loss,NSBH}$.

\subsection{Choice of neutron star equation-of-state} 
\begin{figure}
    \centering
    \includegraphics[width=1.0\linewidth]{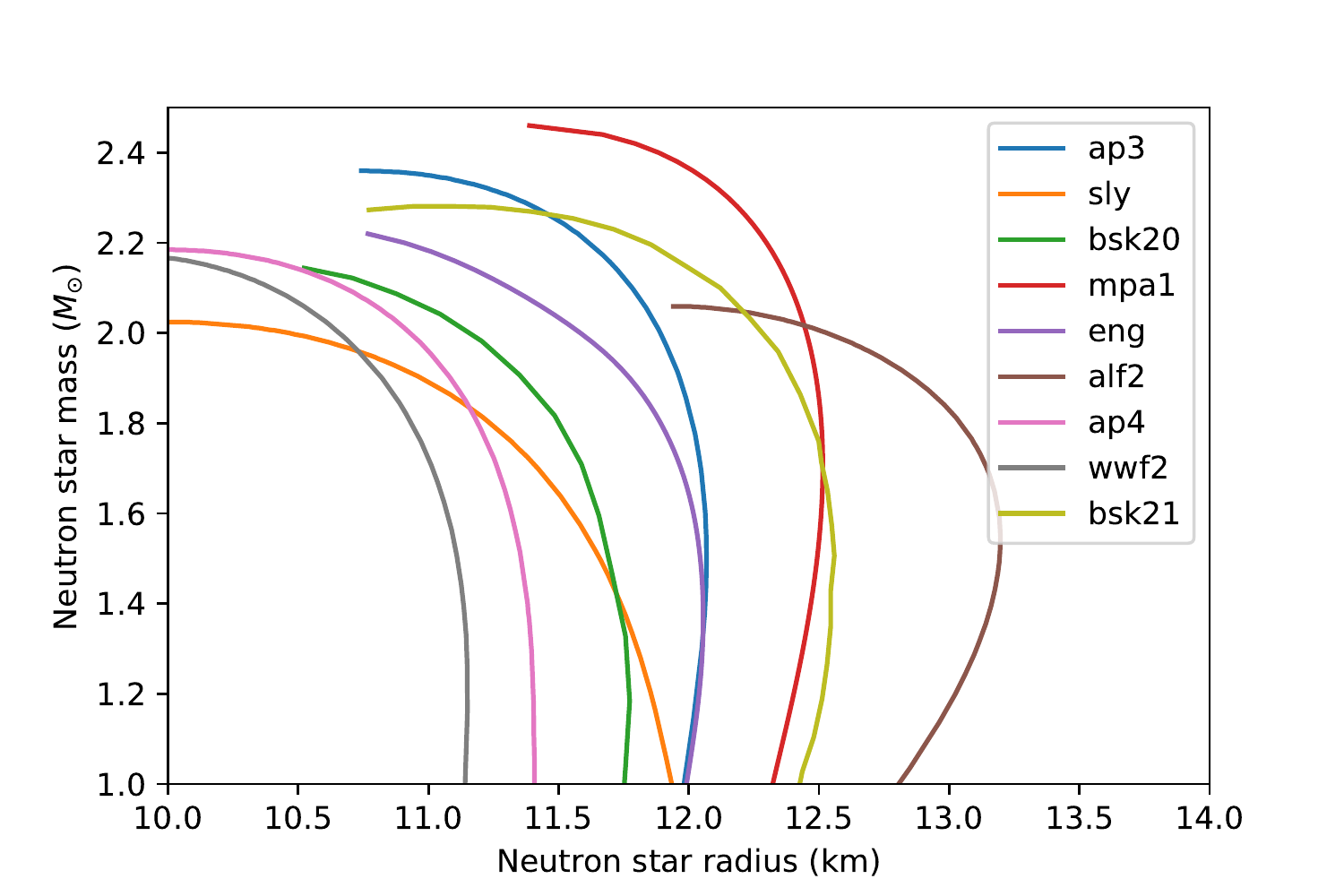}
    \caption{Neutron star EoS provided by the~\href{http://xtreme.as.arizona.edu/NeutronStars/index.php/neutron-star-radii/}{Xtreme Astrophysics Group}~\citep{mr1,mr2,mr3} with mass-radius relations consistent with the combined measurement presented in~\citet{2020ApJ...893L..21R}.}
    \label{fig:mr}
\end{figure}
The amount of ejecta is highly dependent on the compactness of NSs ($C_{\rm NS}\equiv GM/Rc^2$, where $M$ and $R$ are the NS mass and radius, respectively), which is an EoS dependent characteristic. The EoS also affects some of our models setting the minimum mass of the BH mass function (Table~\ref{tab:ratio}).

EoS measurements based on pulsars~\citep{2020NatAs...4...72C,2019ApJ...887L..21R,2019ApJ...887L..24M} and on GW observations of BNSs~\citep{LIGOScientific:2018cki} have been recently combined to yield joint constrains~\citep{2019ApJ...887L..24M,2020ApJ...893L..21R,2021arXiv210506979M,2021arXiv210506980R,2021arXiv210506981R}. 
In order to explore the effect of the choice of EoS, we choose the EoSs yielding mass-radius relations consistent with the 95\% confidence interval of the mass-radius posterior found in ~\citet{2020ApJ...893L..21R}~\footnote{Note that ~\citet{2020ApJ...893L..21R} presented two different model-dependent mass-radius posteriors. To be conservative, we pick all EoSs that are consistent with either of the posteriors.} among the EoSs provided by the~\href{http://xtreme.as.arizona.edu/NeutronStars/index.php/neutron-star-radii/}{Xtreme Astrophysics Group}~\citep{mr1,mr2,mr3}. The selected EoSs are plotted in Figure~\ref{fig:mr}. 

\subsection{Estimation of astrophysical rates}
The merger rate estimates published by the LVK~\citep{o3apop,nsbhdiscovery} assume a model for the mass and spin distribution of compact objects. Since we want to impose different astrophysical models, we don't use the published numbers directly, but instead perform hierarchical Bayesian inference, with the various models detailed in Table~\ref{tab:ratio}. Specifically, we aim at estimating the posterior probability density of the merger rate \err for each population. 
For a population model parametrized by a scale parameter \err (which determines the overall volumetric merger rate of sources) and shape parameters $\Lambda$ (which determines the distribution of sources' parameters), this can be written as~\citep{2002PhRvD..65f3002L,2019MNRAS.486.1086M,2020arXiv200705579V}:

\begin{equation}
p(\err | \Lambda, \{{d}\}) \propto \pi(\err) e^{-N^{s}_{ \uparrow}}\left(\frac{N_{\uparrow}^{s}}{\alpha(\Lambda)}\right)^{N^{\mathrm{tr}}}  \prod_{i=1}^{N^{\mathrm{tr}}} p\left(d \mid \Lambda, \err \right)
\end{equation}

Where $N_{\uparrow}^{s}$ is the number of sources detectable during the observing period, and $N^{\mathrm{tr}}$ is the number of triggers that were actually detected. Both these quantities depend on \err. 
$\alpha(\Lambda)$ is the \emph{fraction} of events that are detectable for a given value of shape parameters, also known as the selection function. We generate samples from this posterior distribution using the \texttt{gwpopulation}~\citep{2019PhRvD.100d3030T} software, and a Jeffrey prior -- $\pi(\err) \propto \err^{-1}$ -- on the rate \err. 

The LVK have not released all the data that would be required to calculate the selection function $\alpha(\Lambda)$ in a way that is identical to what's done for the LVK papers~\footnote{Specifically, the LVK did not release results of simulation campaigns that assess the efficiency of search algorithms in the first observing run for BNS, and in any observing runs for NSBH.}.
Therefore, we rely on a standard procedure in the literature and assess the detection efficiency by using the optimal signal-to-noise ratio (SNR) $\rho$. More specifically, we assume that a binary merger event with true parameters $\vec\theta$ is detectable if its sky and orientation averaged SNR is higher than 8 in a single LIGO detector~\citep{2013ApJ...779...72D}. To account for the diverse sensitivity of the GW detectors in their first three observing runs, we average the detection efficiency of the 3 observing runs as described in Sec 5.2 of~\citet{2020arXiv200705579V}.
To calculate the SNR we use the LIGO-Livingston power spectral density at the time of GW150914 (for the first observing run), GW170817 (for the second observing run) and GW190425 (for the third observing run), as released by~\citet{Abbott:2019ebz}. We do not include spins when calculating the SNR, which has been shown to only impact the detectabilty by less than a few percent~\citep{2018PhRvD..98h3007N}. Finally, we assume that sources are distributed uniformly in comoving volume, and with isotropic sky positions and orientations.
All mass models use the same distribution for the NS in the binary, as described at the beginning of Methods section. The mass spectrum of the BHs is described for each model in Table~\ref{tab:ratio}. We stress that the shape parameters of our models are thus entirely fixed: the only unknown parameter of each population is its overall merger rate \err. 
To infer the rate of BNSs we only use GW170817 and GW190425, whereas when calculating the rate of NSBH we only use GW200105 and GW200115. 

\section{Results} 
We estimate the amount of ejecta for the 100,000 simulated BNSs and NSBHs in each scenario listed in Table~\ref{tab:ratio}, add up the ejecta and scale the sum by the rate ratio of NSBH and BNS, ($R_{\rm NSBH}/R_{\rm BNS}$), to obtain their relative ejecta ratio ($M_{\rm{ej, NSBH}}/M_{\rm{ej, BNS}}$). 
We then estimate the fraction of NSBH ejecta as $M_{\rm{ej, NSBH}}/M_{\rm{ej, Total}}\equiv M_{\rm{ej, NSBH}}/(M_{\rm{ej, BNS}}+M_{\rm{ej, NSBH}})$.
Since these estimations are subject to different sources of uncertainty, we discuss them separately below.

\begin{figure}
    \centering
    \includegraphics[width=1.0\linewidth]{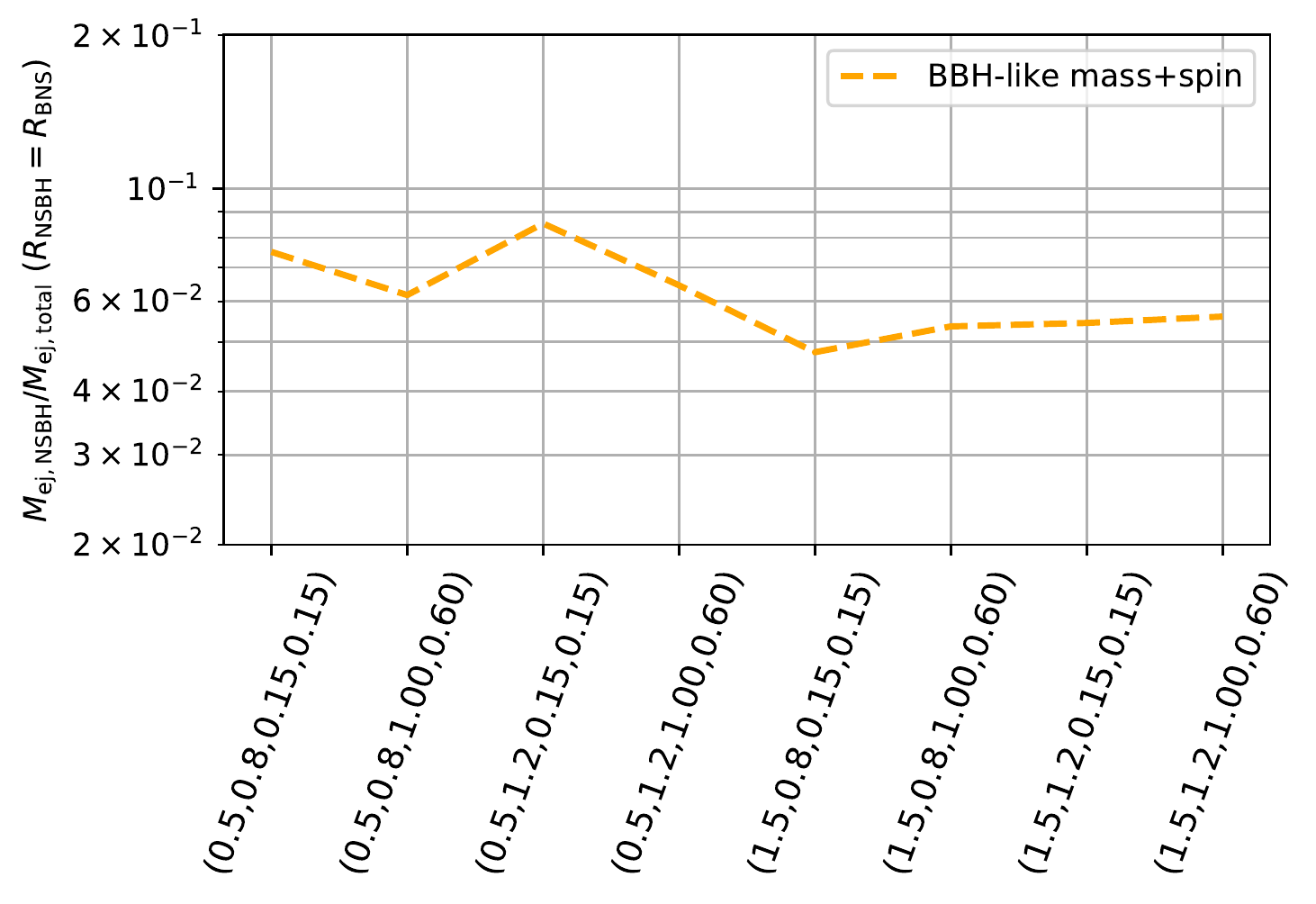}
    \caption{The NSBH ejecta fraction for different ($\alpha_{\rm{dyn,BNS}}$, $\alpha_{\rm{dyn,NSBH}}$, $f_{\rm{loss,BNS}}$ and $f_{\rm{loss,NSBH}}$) [labeled respectively in the horizontal axis] when the BNS and NSBH astrophysical rates are equivalent. We assume the \texttt{BBH-like mass+spin} model and \texttt{ap3} NS EoS in this example. }
    \label{fig:ratio_loss}
\end{figure}

First we show how $\alpha_{\rm{dyn}}$ and $f_{\rm{loss}}$ (Eq.~\ref{eq:tot}) affects the results. We use the \texttt{BBH-like mass+spin} model and the \texttt{ap3} NS EoS.
Figure~\ref{fig:ratio_loss} shows that smaller $\alpha_{\rm{dyn,BNS}}$ and larger $\alpha_{\rm{dyn,NSBH}}$ lead to a larger NSBH fraction as expected. On the other hand, we find that the impact of $f_{\rm{loss,BNS}}$ and $f_{\rm{loss,NSBH}}$ on the NSBH ejecta fraction varies between different NS EoS and the mass and spin models. 

\begin{figure}
    \centering
    \includegraphics[width=1.0\linewidth]{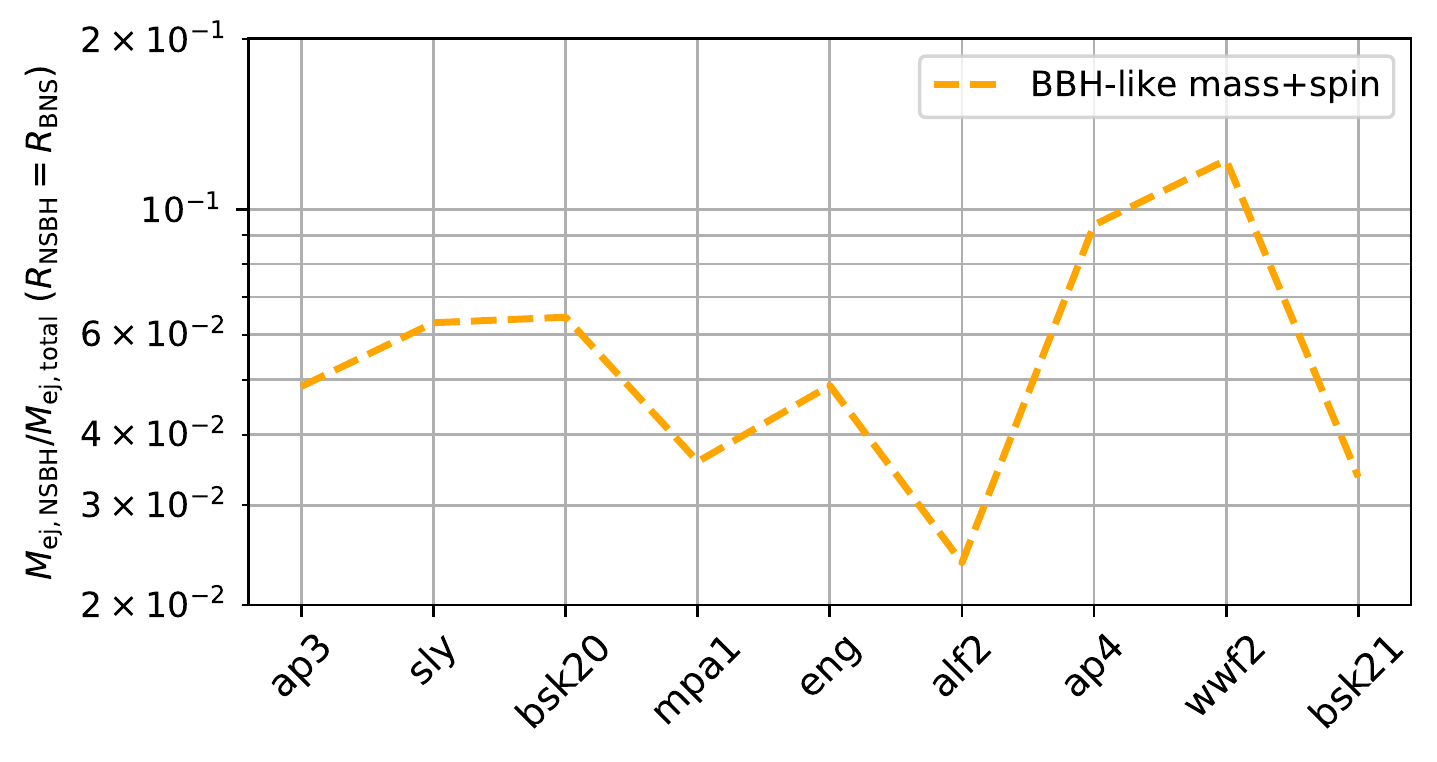}
    \caption{The NSBH ejecta fraction for different EoS shown in Figure~\ref{fig:mr} when the BNS and NSBH astrophysical rates are equivalent.   
    We assume the \texttt{BBH-like mass+spin} model, $\alpha_{\rm{dyn,BNS}}=\alpha_{\rm{dyn,NSBH}}=1$, $f_{\rm{loss,BNS}}=0.8$, and $f_{\rm{loss,NSBH}}=0.4$ in this example. }
    \label{fig:ratio_eos}
\end{figure}

 Next, we show the effect of varying the NS EoS. We use the \texttt{BBH-like mass+spin} model and fix $\alpha_{\rm{dyn,BNS}}=\alpha_{\rm{dyn,NSBH}}=1$, $f_{\rm{loss,BNS}}=0.8$, and $f_{\rm{loss,NSBH}}=0.4$. Figure~\ref{fig:ratio_eos} shows the NSBH ejecta fraction for different EoS shown in Figure~\ref{fig:mr}. 
 In general, stiffer EoS leads to more ejecta for both BNSs and NSBHs. Their impact on the NSBH ejecta fraction depends on the relative ratio between the dynamical ejecta and disk, which vary with $\alpha_{\rm{dyn}}$, $f_{\rm{loss}}$, and the mass and spin models. 

%\begin{figure}
%    \centering
%    \includegraphics[width=1.0\linewidth]{ejectaratio_alluncertainty_rate.pdf}
%    \caption{Fraction of ejecta mass from NSBH as a function of the NSBH and BNS astrophysical rate ratio. The six mass and spin models are %summarized in Table~\ref{tab:ratio}. We assume \texttt{ap3} NS EoS, $\alpha_{\rm{dyn,BNS}}=\alpha_{\rm{dyn,NSBH}}=1$, $f_{\rm{loss,BNS}}=0.8$, %and $f_{\rm{loss,NSBH}}=0.4$ in this figure. In addition, as an example we plot the upper bound for the \texttt{BBH-like mass+aligned spin} %model when the above parameters are allowed to vary (orange band). 
%    We mark the intersection between the upper bound and the 90\% upper limit of rate ratio for the \texttt{BBH-like} mass model. This %intersection represents the upper limit on the NSBH contribution to the production of r-process elements.}
%    \label{fig:ratio}
%\end{figure}

\begin{figure}
    \centering
    \includegraphics[width=1.0\linewidth]{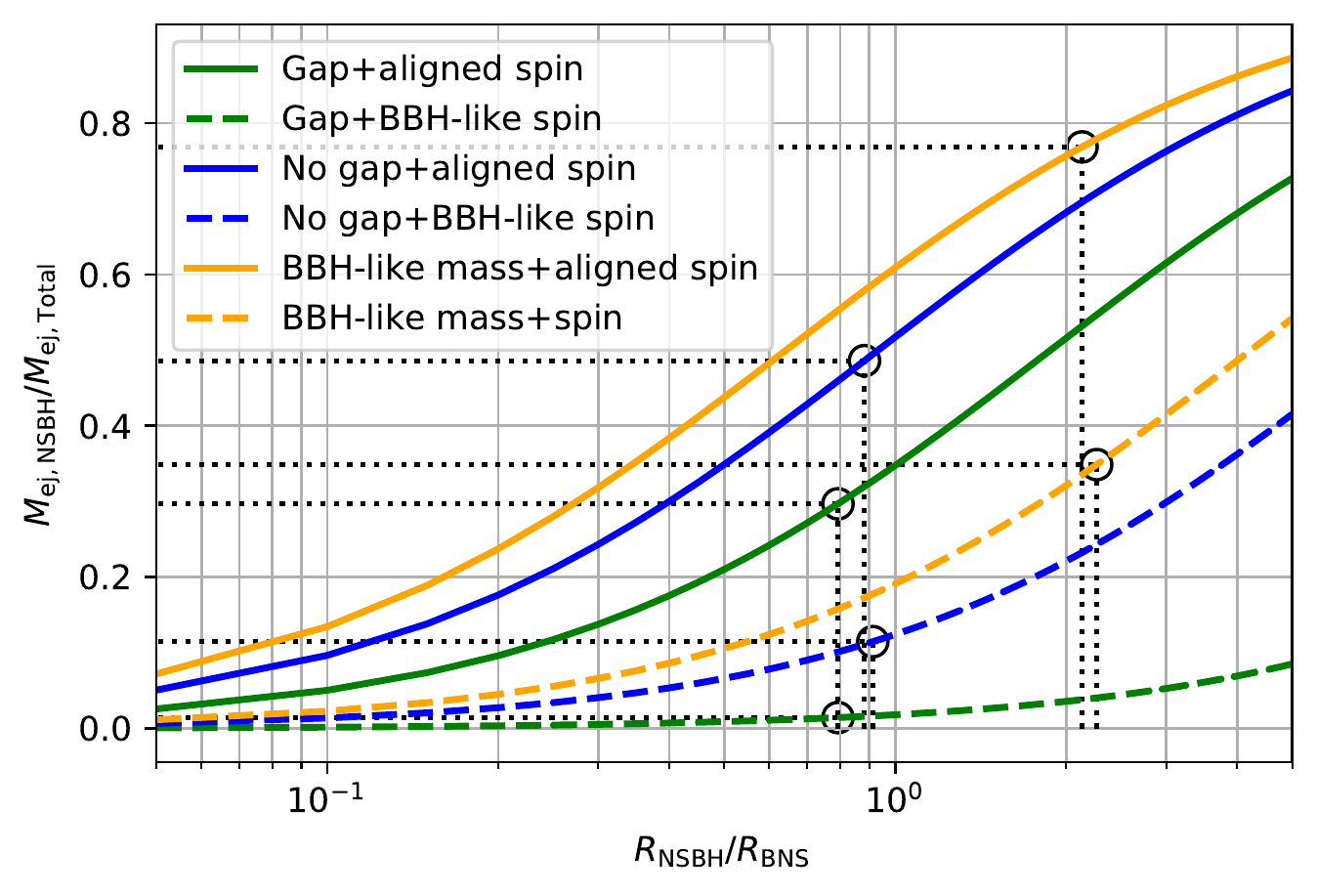}
    \caption{The upper bound of NSBH ejecta fraction as a function of the NSBH and BNS astrophysical rate ratio. The six mass and spin models are summarized in Table~\ref{tab:ratio}. We iterate through different combinations of NS EoS, $\alpha_{\rm{dyn,BNS}}$, $\alpha_{\rm{dyn,NSBH}}$, $f_{\rm{loss,BNS}}$, and $f_{\rm{loss,NSBH}}$, and then use the combinations that lead to the largest NSBH ejecta fractions to set the upper bounds in this figure. 
    We also mark the intersection between the upper bound and the 90\% upper limit of rate ratio for each models. This intersection represents the upper limit on the NSBH contribution to the {binary merger} production of r-process elements. We summarize the upper limit in Table~\ref{tab:ratio}.}
    \label{fig:ratio90}
\end{figure}

In order to determine the overall uncertainty of the ejecta fraction, we combine different sources of error. Since there is no available study on the correlation between the sources of error we discuss above, we conservatively assume they are independent. For each mass and spin model listed in Table~\ref{tab:ratio}, we iterate through different combinations of NS EoS, $\alpha_{\rm{dyn,BNS}}$, $\alpha_{\rm{dyn,NSBH}}$, $f_{\rm{loss,BNS}}$, and $f_{\rm{loss,NSBH}}$. We then use the combination that leads to the largest NSBH ejecta fraction to set the upper bound of our estimations. 
%\sout{In Figure~\ref{fig:ratio} we show the fraction of NSBH ejecta as a function of the rate ratio for each scenario, and as an example, we plot the upper bound for the \texttt{BBH-like mass+aligned spin} model. } 
In Figure~\ref{fig:ratio90} we show the upper bound of the fraction of NSBH ejecta as a function of the rate ratio for each scenario.

We use the LVK observations of BNSs and NSBHs to estimate the astrophysical rate ratio between these two classes of sources. 
%\sout{In Figure~\ref{fig:ratio} we mark the 90\% upper limit of the rate ratio for the \texttt{BBH-like mass} model.} 
In Figure~\ref{fig:ratio90} we mark the 90\% upper limit of the rate ratio for each scenario.
The intersection between the upper bound of NSBH ejecta fraction and the 90\% upper limit of the rate ratio is our constraint on the fraction of NSBH ejecta. Most of the mass ejected from BNSs and NSBHs is expected to become r-process elements, with a small fraction of iron-peak elements that are more likely to be produced in BNSs than NSBHs if the ejecta is less neutron-rich (electron fraction $Y_e\gtrsim 0.3-0.4$, see e.g.~\citet{Korobkin:2012uy,Wanajo2014,Lippuner2015,2021arXiv210511543W}). We thus take the upper limit on the ejected mass fraction to be a good estimate of the limit on NSBH's contribution to the production of r-process elements. The constraints for different scenarios are summarized in the last column of Table~\ref{tab:ratio}.  

NSBHs result in more mass ejected when the mass of BHs is small ($<5M_{\odot}$) or the aligned component of BH spins is large. We find that NSBHs can account for up to 77\% of the r-process-element production. The highest fraction is obtained if there is an excess of 2-3$M_{\odot}$ BHs with spins aligned with orbital angular momentum and uniformly distributed (i.e. for the \texttt{BBH-like mass+aligned spin} model) merging in NSBHs at a high rate. Smaller BH spin magnitudes or larger tilt angles both reduce the aligned component of the spins, leading to less NSBH ejecta. In these cases, the NSBH ejecta fraction decreases to 35\% or less (the \texttt{No gap+BBH-like spin} and \texttt{BBH-like mass+spin} models). These models are also more consistent with the NSBHs observed by LVK so far~\citep{nsbhdiscovery}. If not only BH spins are small, but there also is a dearth of low-mass ($<5M_{\odot}$) BHs in NSBHs, then BNSs produce virtually all of the r-process elements from compact binary mergers. We therefore do not explore models with even larger BH mass gap or smaller BH spins since they would lead to even less NSBH ejecta. 

\section{Discussion} 
In this paper we have combined GW detections and pulsar observations to place {constraints on the relative} contribution of BNSs and NSBHs to the production of r-process elements. NSBHs contribution can be as high as 77\%, if BHs in NSBHs can have low mass (2-3$M_{\odot}$) and large spins aligned with the orbital angular momentum. {However, this low-mass and high-spin BHs scenario seems disfavoured by GW observations of NSBHs. I}f most black holes have masses in excess of $\sim 5~M_{\odot}$ and/or small spins, BNS will contribute the nearly entirety of r-process elements from compact binary mergers. Different EoS and ejecta models can lower the upper limits by up to 30\%, which is smaller than the differences arising from different mass and spin models.

The relative contribution of BNSs and NSBHs to the production of heavy elements depends thus significantly on the distribution of black holes masses and spins in binaries. These will be better measured by upcoming LVK observations~\citep{Aasi:2013wya}, leading to more precise estimates of the ejecta ratio. In addition, current LVK observations of BNSs and NSBHs are local ($z<0.2$), thus our inference constrains the {binary merger} production of the r-process elements over the past 2.5 billion years. Future GW detections at higher redshifts will enable measuring the evolution of BNS and NSBH astrophysical rates and therefore their redshift-dependent contribution to the production of heavy elements.

Although our analysis yields an estimate of the BNS ejecta fraction ($\equiv 1-M_{\rm{ej, NSBH}}/M_{\rm{ej, Total}}$), we do not present a lower limit on the BNS ejecta since other astrophysical channels could still produce some of the heavy elements. The constraints we have obtained are mainly limited by the number of available numerical simulations. Whenever different options existed, we have conservatively take those that yielded the largest error bars. A more extensive coverage of the physical parameter space by numerical simulations of BNS and NSBH mergers will thus reduce the uncertainty in the estimation of the mass ejecta. 

This work demonstrates the potential of combining electromagnetic and gravitational-wave observations of compact objects. It also underlines the need for more extensive numerical simulations of the mass ejecta from compact objects. As advanced gravitational-wave detectors will measure with more precision the mass and spin distribution of compact objects, whilst gravitational-wave and pulsar observations will yield improved constraints on the NS EoS, we will be able to significantly refine the constraints presented here in the months and years to come.

\prlsec{Acknowledgments} We thank Katerina Chatziioannou and Will Farr for help using the data release associated with~\citet{2020RNAAS...4...65F}. We thank Justin Alsing, Emanuele Berti, Zoheyr Doctor, Anna Frebel, and Brian Metzger for valuable comments and suggestions.
HYC is supported by NASA through NASA Hubble Fellowship grants No.\ HST-HF2-51452.001-A awarded by the Space Telescope Science Institute, which is operated by the Association of Universities for Research in Astronomy, Inc., for NASA, under contract NAS5-26555. 
S.V. acknowledges support of the National Science Foundation and the LIGO Laboratory. LIGO was constructed by the California Institute of Technology and Massachusetts Institute of Technology with funding from the National Science Foundation and operates under Cooperative Agreement No. PHY-1764464.
F.F. gratefully acknowledges support from the NSF through grant PHY-1806278, from the DOE through grant DE-SC0020435, and from NASA through grant 80NSSC18K0565. This research has made use of data, software and/or web tools obtained from the Gravitational Wave Open Science Center (https://www.gw-openscience.org/ ), a service of LIGO Laboratory, the LIGO Scientific Collaboration and the Virgo Collaboration. LIGO Laboratory and Advanced LIGO are funded by the United States National Science Foundation (NSF) as well as the Science and Technology Facilities Council (STFC) of the United Kingdom, the Max-Planck-Society (MPS), and the State of Niedersachsen/Germany for support of the construction of Advanced LIGO and construction and operation of the GEO600 detector. Additional support for Advanced LIGO was provided by the Australian Research Council. Virgo is funded, through the European Gravitational Observatory (EGO), by the French Centre National de Recherche Scientifique (CNRS), the Italian Istituto Nazionale di Fisica Nucleare (INFN) and the Dutch Nikhef, with contributions by institutions from Belgium, Germany, Greece, Hungary, Ireland, Japan, Monaco, Poland, Portugal, Spain.

\bibliography{ref}% Produces the bibliography via BibTeX.

\end{document}